\documentclass[amstex,aps,preprint,eqsecnum]{revtex4}

\begin{document}

\title{Fluctuations and Noise: A General Model with Applications}

\author{R. F. O'Connell{\footnote{E-mail: oconnell@phys.lsu.edu \\
Phone: (225) 578-6848 Fax: (225) 578-5855
\\ }}}

\affiliation{Department of Physics and Astronomy, Louisiana State
University, \\ Baton Rouge, Louisiana  70803-4001, USA}

\date{\today}

\begin{abstract} 
A wide variety of dissipative and fluctuation problems involving a
quantum system in a heat bath can be described by the
independent-oscillator (IO) model Hamiltonian.  Using Heisenberg
equations of motion, this leads to a generalized quantum Langevin
equation (QLE) for the quantum system involving two quantities
which encapsulate the properties of the heat bath.  Applications
include:  atomic energy shifts in a blackbody radiation heat bath;
solution of the problem of runaway solutions in QED; electrical
circuits (resistively shunted Josephson barrier, microscopic
tunnel junction, etc.); conductivity calculations (since the QLE
gives a natural separation between dissipative and fluctuation
forces); dissipative quantum tunneling; noise effects in
gravitational wave detectors; anomalous diffusion; strongly driven
quantum systems; decoherence phenomena; analysis of Unruh
radiation and entropy for a dissipative system.
\\
\\
\\
\\
\\
\\
\\
\noindent {\bf{Keywords:}}  Noise, fluctuations, dissipation,
decoherence, Langevin equation, blackbody radiation, Unruh
radiation
\end{abstract}

\maketitle

\newpage

\section{INTRODUCTION}

Noise is due to rapid fluctuations in the average of physical
quantities.  There are many types of classical noise (thermal,
...) but, in addition, there exits intrinsic quantum noise.  A
wide variety of noise problems can be described in a universal
manner by means of a generalized quantum Langevin equation (QLE)
\cite{ford}. This equation goes beyond the classical Langevin
equation \cite{langevin} by incorporating a potential, both
quantum and non-Markovian effects, the quantum
fluctuation-dissipation theorem, \cite{callen} as well as the
presence of a time-dependent external force (thus allowing a
description of the evolution of an irreversible system).  Our
purpose here is to outline how one obtains the QLE by starting
with a very general microscopic model.  Next, we consider a
variety of applications.  Before proceeding, we will first give a
brief history of different approaches to fluctuation and
dissipative processes.

As distinct from a dynamical system, a heat bath (reservoir)
describes a subsystem with an infinite number of degrees of
freedom which enable the system (i.e., the particle interacting
with the heat bath) to relax in the course of time to a unique
equilibrium state.  A commonly used model of a heat bath is an
infinite number of oscillators and the freedom of choice of the
relevant masses and frequencies leads to a surprising diversity in
the physical systems described, examples being blackbody radiation
(BBR) and phonon heat baths and impurities in a metal (in the
limit of the oscillator frequencies going to zero).

The beginning of the subject is generally regarded as having
started with the observations of Robert Brown, a Scottish botanist
who observed the random motion of pollen grains immersed in a
fluid \cite{brown}.  No external forces were present and the
temperature $T$ was room temperature.  Later work showed that the
very irregular motion occurred in any suspension of tiny particles
in a liquid medium.  The explanation of these phenomena was given
in a series of papers by Einstein \cite{einstein} from 1905 to
1908.  Coupled with the experimental work of Perrin \cite{perrin}
this served to establish the atomic theory of matter since the
irregular motion was clearly identified as being due to collisions
with the molecules in the liquid.  The term "Brownian motion" is
now used in a generic sense to denote random motion and it covers
a wide spectrum of phenomena from the motion of very fine
particles suspended in gas to the motion of electrons in a BBR
heat bath.  \emph{Einstein's} explanation of Brownian motion used a
\textit{discrete time} approach.  In particular, his result that
the diffusion constant
$D$ (which is defined as one-half of the rate of change with time
$t$ of the mean-square displacement in the limit of large $t$ is
proportional to $T/{\gamma}$, where $T$ is the temperature and
${\gamma^{-1}}$ is the collision rate, is the first example of a
fluctuation - dissipation (FD) theorem.

Shortly after the work of Einstein, Langevin \cite{langevin}
presented an entirely new approach to the subject which, in the
words of Chandrasekhar \cite{chandrasekhar}, constitutes the
"modern" approach to this and other such problems.  The essence of
\emph{Langevin's} approach is a
\emph{continuous time} approach implemented by adoption of a
\emph{stochastic differential equation}, i.e., an equation for
quantities which are \emph{random} in nature.  In other words,
Langevin provided an elegant solution to the problem of
generalizing a dynamical equation to a probabilistic equation. 
This was to be the start of a major new field of study with
widespread implications in physics, chemistry, biology, and many
other fields.  The approach of Langevin was phenomenological in
nature but its essential correctness has been verified by various
microscopic studies.  An essential feature of his approach was to
separate the total force acting on a particle due to its
environment into two parts: a frictional force and a fluctuation
(random) force.  These terms are very different in nature:  The
fluctuation term is basically microscopic in nature and has a time
scale determined by the mean time between collisions whereas the
time scale of the frictional force is proportional to the
self-diffusion constant and is much larger.  

Another example of a fluctuation-dissipation theorem was provided
by Nyquist \cite{nyquist} who showed that the random fluctuations
in voltage and current across a resistor measured by Johnson
\cite{johnson} are determined by its impedance (the famous
Johnson-Nyquist noise in electrical circuits).  All of the
aforementioned work was classical in nature, but in 1951, Callen
and Welton \cite{callen} presented their celebrated work on a
quantum formulation of the FD theorem.  Since a major shortcoming
of the Langevin equation is its phenomenological nature, it is
clearly desirable to obtain such an equation from
{\underline{microscopic}} considerations.  Such a theory was given
by Zwanzig \cite{zwanzig} who considered the particle of interest
to interact with an environment consisting of an infinite number
of oscillators and obtained the usual classical high-temperature
Brownian motion results.  Turning to microscopic quantum
dissipative phenomena, pioneering approaches to developing a
quantum Langevin equation for particular problems appear in the
articles of Senitzky \cite{senitzky} and Lax \cite{lax}.  However,
these articles were based on a Markovian approximation and,
unfortunately, the work of Senitzky was marred by serious errors,
as pointed out by Li et al.\cite{li}  Perhaps the most influential
of the earlier articles is perhaps the oft-quoted article of Ford
et al. \cite{ford65} since it was the first article in which the
correct formulation of the quantum Langevin equation was
indicated.  Other important contributions include the work of Mori
\cite{mori}, as well as Benguria and Kac
\cite{benguria} and Ford and Kac \cite{ford87}, leading up to the
work of Ford, Lewis, and O'Connell (FLO) \cite{ford} who, in an
article entitled "Quantum Langevin Equation", gave a detailed
discussion of the problem.  In particular, these authors presented
the general form of the QLE consistent with fundamental physical
requirements, in particular, causality and the second law of
thermodynamics.  Other approaches to dissipative problems include
Kubo's linear response theory \cite{kubo},\cite{kubo66} and the
Feynman-Vernon use of path integrals which has been condiderably
extended by Leggett \cite{caldeira} and co-workers as well as many
German investigators \cite{grabert},\cite{weiss} among others. 
However, the QLE approach is our method of choice since not only
is it at least on a par technically with all the other approaches
but we also consider it to be more physically appealing and
simpler to execute.

Sec. 2 is devoted to fundamentals.  In particular, we show how the
QLE can be obtained from a very general and basic Hamiltonian
after which we proceed to show how this leads to results for
important quantum commutators and correlations which form the
basis for the derivation of observable quantities.  The QLE
derived in Ref. 1 pertains to a stationary process, in the sense
that correlations, probability distributions, etc. for the
dynamical variable $x$ are invariant under time-translation $(t
\rightarrow t+t_{0})$. It is this equation which is the basis of
most of the applications which we discuss below.  There is also a
QLE for the initial value problem \cite{ford87} which has been
used to obtain the most general master with explicit expression
for the associated time-dependent coefficients, leading in turn to
an explicit exact solution \cite{ford01}.  Sec. 3 discusses the
Ohmic heat bath model and various applications and similarly for
the radiation heat bath model in Sec. 4.  Driven Systems are
discussed in Sec. 5.
\\

\section{GENERALIZED QUANTUM LANGEVIN EQUATION}

The QLE is a Heisenberg equation of motion for the coordinate
operator $x$ of a quantum particle of mass $m$ moving in a
one-dimensional potential $V(x)$ and linearly coupled to a passive
heat bath at temperature $T$.  Whereas it has a very general form,
as we pointed out in Ref. 1, it can be realized with a simple and
convenient model, viz., the independent-oscillator (IO) model. 
The Hamiltonian of the IO system is
\begin{equation}
H=\frac{p^{2}}{2m}+V(x)+\sum_{j}\left(\frac{p^{2}_{j}}{2m_{j}}+\frac{1}{2}m_{j}\omega^{2}_{j}(q_{j}-x)^{2}
\right)-xf(t). \label{dn22}
\end{equation} Here $m$ is the mass of the quantum particle while
$m_{j}$ and
$\omega_{j}$ refer to the mass and frequency of heat-bath
oscillator
$j$.  In addition, $x$ and $p$ are the coordinate and momentum
operators for the quantum particle and $q_{j}$ and $p_{j}$ are the
corresponding quantities for the heat-bath oscillators.  Also
$f(t)$ is a c-number external force.  The infinity of choices for
the $m_{j}$ and $\omega_{j}$ give this model its great
generality.  In particular, it can describe nonrelativistic
quantum electrodynamics (QED), the Schwabl-Thirring model, the
Ford-Kac-Mazur model, and the Lamb model \cite{ford}.

Use of the Heisenberg equations of motion leads to the QLE 
\begin{equation} m \ddot{x} +\int^{t}_{-\infty}dt^{\prime}\mu
(t-t^{\prime})\dot{x}(t^{\prime})+V^{\prime}(x)=F(t)+f(t),
\label{dn21}
\end{equation}  where $V^{\prime}(x)=dV(x)/dx$ is the negative of
the time-independent external force and $\mu (t)$ is the so-called
memory function.  $F(t)$ is the random (fluctuation or noise)
operator force with mean $\langle F(t)\rangle =0$ and
$f(t)$ is a $c$-number external force (due to an electric field,
for instance).  In addition, it should be strongly emphasized that
"-- the description is more general than the language --"
\cite{ford} in that
$x(t)$ can be a generalized displacement operator (such as the
phase difference of the superconducting wave  function across a
Josephson junction).

Thus, the coupling with the heat bath is described by two terms: 
an operator-valued random force $F(t)$ with mean zero, and a mean
force characterized by a memory function $\mu(t)$.  Explicitly,
\begin{equation}
\mu(t)=\sum_{j}m_{j}\omega^{2}_{j}\cos(\omega_{j}t)\theta(t),
\label{dn23}
\end{equation}, with $\theta (t)$ the Heaviside step function. 
Also
\begin{equation} F(t)=\sum_{j}m_{j}\omega^{2}_{j}q^{h}_{j}(t),
\label{dn24}
\end{equation} is a fluctuating operator force with mean $\langle
F(t)\rangle =0$, where $q^{h}(t)$ denotes the general solution of
the homogeneous equation for the heat-bath oscillators
(corresponding to no interaction).  These results were used to
obtain the results for the (symmetric) autocorrelation and
commutator of $F(t)$, viz.,
\begin{eqnarray} C_{FF}(t-t^{\prime})&=&\frac{1}{2}\langle
F(t)F(t^{\prime})+F(t^{\prime})F(t)\rangle
\nonumber \\ &=&
\frac{1}{\pi}\int^{\infty}_{0}d\omega
Re[\tilde{\mu}(\omega+i0^{+})]\hbar\omega\coth
(\hbar\omega/2kT)\cos[\omega(t-t^{\prime})], \label{dn25}
\end{eqnarray}
\begin{eqnarray} [F(t),F(t^{\prime})] &=&
\frac{2\hbar}{i\pi}\int^{\infty}_{0}d\omega Re\{\tilde{\mu}(\omega
+i0^{+})\}\omega\sin\omega(t-t^{\prime}). \label{dn26}
\end{eqnarray} We note that both of these results are independent
of $V(x)$ and $f(t)$. Here
$\tilde{\mu}(z)$ is the Fourier transform of the memory function:
\begin{equation}
\tilde{\mu}(z)=\int^{\infty}_{0}dt\mu (t)e^{izt}. \label{dn27}
\end{equation}

Kubo \cite{kubo},\cite{kubo66} refers to (2.5) as the second
fluctuation-dissipation theorem and we note that it can be written
down explicitly once the QLE is obtained.  Also, its evaluation
requires only knowledge of Re$\tilde{\mu}(\omega)$.  On the other
hand, the first fluctuation-dissipation theorem is an equation
involving the autocorrelation of $x(t)$ and its explicit
evaluation requires a knowledge of the generalized susceptibility
$\alpha(\omega)$ (to be defined below) which is equivalent to
knowing the solution to the QLE and also requires knowledge of
both Re$\tilde{\mu}(\omega)$ and Im$\tilde{\mu}(\omega)$.  This
solution is readily obtained when
$V(x)=0$, corresponding to the original Brownian motion problem
\cite{langevin}.  As shown in Ref. \cite{ford85}, a solution is
also possible in the case of an oscillator.  Taking
$V(x)=\frac{1}{2}Kx^{2}=\frac{1}{2}m\omega^{2}_{0}x^{2}$, these
authors obtained the solution of the  Langevin equation (2.2) in
the form
\begin{equation} x(t)=\int_{-\infty }^{t}dt^{\prime }G(t-t^{\prime
})\{F(t^{\prime })+f(t^{\prime})\},
\label{dn28}
\end{equation} where $G(t)$, the Green function, is given by
 \begin{equation} G(t)=\frac{1}{2\pi }\int_{-\infty }^{\infty
}d\omega
\alpha (\omega +i0^{+})e^{-i\omega t},  \label{dn29}
\end{equation} with $\alpha (z)$ the familiar response function
\begin{equation}
\alpha (z)=\frac{1}{-mz^{2}-iz\tilde{\mu}(z)+K}.  \label{dn210}
\end{equation} Also, taking the Fourier transform of (\ref{dn28}),
we obtain
\begin{equation}
\tilde{x}(\omega)=\alpha(\omega)\{\tilde{F}(\omega)+\tilde{f}(\omega)\},
\label{dn214}
\end{equation} where the superposed tilde is used to denote the
Fourier transform.  Thus, $\tilde{x}(\omega)$ is the Fourier
transform of the operator $x(t)$:
\begin{equation}
\tilde{x}(\omega)=\int^{\infty}_{-\infty}dtx(t)e^{i\omega t}.
\label{dn215}
\end{equation} It is also useful to note that the commutator,
which is temperature independent, is given by the formula
\cite{ford89}
\begin{equation}
\lbrack x(t_{1}),x(t_{1}+t)]=\frac{2i\hbar }{\pi }\int_{0}^{\infty
}d\omega \mathrm{Im}\{\alpha (\omega +i0^{+})\}\sin
\omega t. \label{dn216}
\end{equation}

Whereas the autocorrelation and commutator of $F(t)$ are
independent of $V(x)$ and $f(t)$, this is not so in the case of
$x(t)$, as is obvious from (2.8) and (2.10).  However, since most
of our applications pertain to the case $f(t)=0$, we will confine
ourselves for the moment to this case and discuss later
generalizations which are necessary when $f(t)\ne0$.  Thus, in
particular, when $f(t)=0$ 
\begin{eqnarray} C(t-t^{\prime})&&\equiv\frac{1}{2}\langle
x(t)x(t^{\prime })+x(t^{\prime })x(t)\rangle \nonumber \\ 
&&{}=\frac{\hbar }{\pi }%
\int_{0}^{\infty }d\omega {\rm Im}\{\alpha (\omega +i0^{+})\}\coth
\frac{
\hbar \omega }{2kT}\cos \omega (t-t^{\prime }).  \label{dn212}
\end{eqnarray} We note that it is a function only of the
time-difference $t-t^{\prime }$.

This is referred to by Kubo \cite{kubo},\cite{kubo66} as the
fluctuation-dissipation theorem of the first kind.  Next, from
(2.10), we see that 
\begin{equation} Im~~
\alpha(\omega)=\omega|\alpha(\omega)|^{2}~~Re~\tilde{\mu}(\omega).
\end{equation} Thus, (2.14) may be expressed in the form
\begin{equation} C(t)=\frac{\hbar }{\pi }%
\int_{0}^{\infty
}d\omega~~\omega|\alpha(\omega)|^{2}~~Re~\tilde{\mu}(\omega)~~\coth\left(\frac{\hbar\omega}{{2}\kappa{T}}\right)
\cos~\omega{t}.
\end{equation} The mean square displacement of the quantum
particle in a dissipative environment,
$s(t)$ say, plays a key role in many of our subsequent
discussions.  In particular, it determines the diffusion time
through a  system of interest.  Thus using (\ref{dn212}), we obtain
\begin{eqnarray} s(t) &\equiv &\left\langle [x_{{\rm s}}(t)-x_{{\rm
s}}(0)]^{2}\right\rangle 
\nonumber \\ &=& 2\left\{C(0)-C(t)\right\} \nonumber \\
&=&\frac{2\hbar }{\pi }\int_{0}^{\infty }d\omega {\rm Im}\{\alpha
(\omega +i0^{+})\}\coth \frac{\hbar \omega }{2kT}(1-\cos \omega
t). 
\label{dn31}
\end{eqnarray} 

For specific applications, the key ingredient to be identified is
the nature of the heat bath which is simply characterized by
$Re~\tilde{\mu}(\omega+i0^{+})$, the spectral distribution of the
memory fuction, which is a positive real function \cite{ford} and
given explicitly by the relation
\begin{equation}
Re~[\tilde{\mu}(\omega+i0^{+})]=\frac{\pi}{2}\sum_{j}m_{j}\omega^{2}_{j}[\delta(\omega-\omega_{j})+
\delta(\omega+\omega_{j})].
\end{equation} Next, we summarize how the above tools have been
used to obtain thermodynamic quantities, especially results which
have been obtained only by the QLE method.  The key quantity
calculated \cite{ford85} is the free energy \emph{F}, which is a
thermodynamic potential from which other thermodynamic functions
can be obtained by differentiation.  

The system of an oscillator coupled to a heat bath in thermal
equilibrium at temperature \emph{T} has a well-defined free
energy.  The free energy ascribed to the oscillator, \emph{F(T)},
is given by the free energy of the system minus the free energy of
the heat bath in the absence of the oscillator.  This calculation
was carried out by two different methods
\cite{ford85},\cite{ford88} leading to the "remarkable formula":
\begin{eqnarray}
F(T)=\frac{1}{\pi}\int_{0}^{\infty}d\omega{f}(\omega,T)
\times{Im}\left\{\frac{d\log\alpha(\omega+i{0}^+)}{d\omega}\right\},
\end{eqnarray} where $f(\omega,T)$ is the free energy of a single
oscillator of frequency $\omega$, given by
\begin{equation}
f(\omega,T)=\kappa{T}\log\left[1-\exp\left(\frac{-\hbar\omega}{\kappa{T}}\right)\right]. 
\end{equation}  Here the zero-point contribution
$(\frac{\hbar\omega}{2})$ has been omitted. In particular, this
result was used to obtain the energy and the entrophy ascribed to
the osillator.

Finally, we note that magnetic field effects on all of the above
quantities have been calculated
\cite{li90},\cite{li96}.  In particular, it was shown that the
effect of an arbitrary heat bath on Landau diamagnetism is always
such as to reduce the magnitude of the magnetic moment without
changing its diamagnetic character \cite{li96}.

Thus, we now have all the tools necessary for applications.  For
each particular application, one must first decide on the correct
specification of the relevant heat bath. The case of constant
friction, the so-called Ohmic model, is of special interest since
it is the simplest model but yet gives a good description of many
physical systems.  A generalization of this model, involving an
additional parameter, is the single relaxation time model
\cite{ford01}.  Also of interest is the Debye model \cite{ludu}. 
The blackbody radiation (BBR) heat bath model
\cite{ford},\cite{ford85} is of fundamental interest since the
relevant Hamiltonian \emph{H} in this case is the universally
accepted \emph{H} of quantum electrodynamics.  Thus, we refer to
the BBR model as "the rosetta-stone of heat bath models" since it
leads to readily-checked predictions and it is the basis of many
of the applications which we consider below.  Hence, in Sec. 3, we
consider the Ohmic model and various applications and we do
likewise for the BBR model in Sec. 4.

\section{OHMIC HEAT BATH MODEL AND APPLICATIONS}

The Ohmic model is defined by the choice

\begin{equation}
\mu(t)=\zeta\delta(t), ~~~t>0,
\end{equation} so that
\begin{equation} ~\tilde{\mu}(\omega)=\zeta=m\gamma=constant.
\end{equation}  Also, it follows that
\begin{equation} m(\ddot{x}+\gamma\dot{x}+\omega^{2}_{0})=F(t).
\end{equation}  This corresponds to the original form of the
Langevin equation except that now the classical quantities $x(t)$
and $F(t)$ are operators and we have also included an oscillator
potential.  Since the past motion does not appear, one says there
is no memeory.  On the other hand, the quantum-mechanical process
is not Markovian since $C_{FF}(t-t^{\prime})$ is not proportional
to $C_{FF}(t-t^{\prime})$ a $\delta(t-t^{\prime})$ except in the
classical limit where
$\hbar\rightarrow0$  \cite{ford}.  Using (3.3) in (2.10) leads to
the result
\begin{equation}
\alpha(\omega)=\{m[-\omega^{2}+\omega^{2}_{0}-i\omega\gamma]\}^{-1},
\end{equation} from which it follows, using (2.9), that the green
fucition is given by
\begin{equation}
G(t)=e^{-{(\frac{\gamma{t}}{2})}}~~~\frac{\sin\omega_{1}t}{m\omega_{1}},
\end{equation} where
\begin{equation}
\omega_{1}=\{\omega^{2}_{0}-(\frac{\gamma}{2})^{2}\}^{\frac{1}{2}}.
\end{equation} Also, from (2.16), we obtain
\begin{equation}
C(t)=\frac{\hbar\gamma}{2\pi{m}}\int_{-\infty}^{\infty}~~~\frac{\omega\coth(\frac{\hbar\omega}{2kT})}
{(\omega^{2}_{0}-\omega^{2})^{2}+\gamma^{2}\omega^{2}}.
\end{equation} We now consider various applications.

\subsection{Brownian Motion}

This is the original problem of diffusion of a free particle
through a medium characterized by a dissipative decay rate
$\gamma$.  Since here $\omega_{0}=0$, the results given above
reduce to 

\begin{eqnarray}
\alpha(\omega)&=&\{m[-\omega^{2}-i\omega\gamma]\}^{-1},\\
G(t)&=&\frac{1-e^{-\gamma{t}}}{m\gamma},\\ C(t)
&=&\frac{\hbar\gamma}{2\pi{m}}\int_{-\infty}^{\infty}d\omega~~~\frac{\coth(\frac{\hbar\omega}{2kT})}
{\omega(\omega^{2}+\gamma^{2})}~~~e^{-i\omega{t}}.
\end{eqnarray} Thus, (2.17) gives
\begin{equation}
s(t)=\frac{\hbar\gamma}{\pi{m}}\int_{-\infty}^{\infty}d\omega~~~\frac{(1-\cos\omega{t})}{\omega
(\omega^{2}+\gamma^{2})}~~~\coth\left(\frac{\hbar\omega}{2kT}\right).
\end{equation}  In the classical high temperature limit
(corresponding to the Brown experiment), 
$\coth\left(\frac{\hbar\omega}{2kT}\right)\longrightarrow(\frac{2kT}{\hbar\omega})$,
so that
\begin{equation} s(t)=\frac{2kT}{m\gamma^{2}}\left\{e^{-\gamma
t}-1+\gamma t\right\}, ~~~ kT>>\hbar\gamma. \label{dn34}
\end{equation} For long times
\begin{equation} s(t)\rightarrow \frac{2kT}{m\gamma} t,~~~\gamma
t>>1,
\label{dn35}
\end{equation} which is the familiar Einstein relation from
Brownian motion theory.  However, for short times (which are more
characteristic of decoherence decay times)
\begin{equation} s(t)\rightarrow\frac{kT}{m}t^{2},~~~\gamma t<<1,
\label{dn36}
\end{equation} independent of $\gamma$. Thus, in this simple case,
the familiar Einstein diffusion coefficient 
\begin{equation} D\equiv \frac{1}{2}\dot{s}(t)=\frac{kT}{m\gamma},
\label{dn37}
\end{equation} is obtained.

In particular, the situation is very different for low temperatures
$(kT<<\hbar\gamma)$, in which case the main contribution is from
the zero-point $(T=0)$ oscillations of the electromagnetic field. 
This calculation has recently been carried out \cite{ford03}
leading to the result
\begin{equation}
s(t)\cong-\frac{\hbar\zeta}{\pi{m}^{2}}t^{2}\left\{\log\frac{\zeta{t}}{m}+\gamma_{E}-\frac{3}{2}
\right\},
\end{equation}where $\gamma_{E}=0.577215665$ is Euler's constant.

\subsection{Noise in gravitational wave detector suspension
systems}

Crucial to the detection of gravitational waves is the mechanical
system which is used to measure the relative displacements of
suspended mirrors [interferometric experiments, such as the Laser
Interferometric Gravitational Wave Observatory (LIGO) or the
displacement of a resonant mass antenna (bar experiments)].  Here,
we present a general framework for describing noise effects in
detectors systems.  Our results apply to a very general
dissipative environment but for definiteness we will concentrate
mostly on the LIGO-like test mass suspensions and on bar detectors.

The crucial quantity to be calculated is the ensemble average of
the square of the displacement due to noise,
$\langle{x}^{2}(t)\rangle$, which is simply equal to C(0), as is
clear from (2.14).  Hence, from (2.16),
\begin{eqnarray}
\langle{x}^{2}(t)\rangle&=&\frac{\hbar}{\pi}\int_{0}^{\infty}d\omega\omega\mid\alpha(\omega)\mid^{2}
Re\widetilde{\mu}(\omega)\coth\left(\frac{\hbar\omega}{2kT}\right)\equiv\int_{0}^{\infty}P(\omega)d\omega,
\end{eqnarray} where
\begin{equation}
P(\omega)=\frac{\hbar}{\pi}\omega\mid\alpha(\omega)\mid^{2}Re\widetilde{\mu}(\omega)\coth\left(\frac{\hbar
\omega}{2kT}\right),
\end{equation} is the power spectrum of the coordinate
fluctuations.

In the case of resonant bar detectors, $\widetilde{\mu}(\omega)$
is taken to be $m\gamma$, where
$\gamma$ is a constant.  Thus, in particular, if we take the
weak-coupling limit $\omega_{0}
\ll\gamma$ and substitute (3.4) into (3.17), we immediately find
the well-known high-temperature result
$\langle{x}^{2}(t)\rangle=\frac{kT}{m\omega^{2}_{0}}$ and the
zero-temperature result
$\langle{x}^{2}(t)\rangle=(\frac{\hbar}{2m\omega_{0}})$
\cite{roc}.  However, in the case of nonresonant LIGO detectors,
which are responsive to a range of frequencies, the frequency
dependence of $\widetilde{\mu}(\omega)$ is essential.  In
practice, because of the complexity of the detector system, the
practical procedure will be to fit the experimental results by some
$\widetilde{\mu}(\omega)$.  This has the advantage that we know a
lot about the properties of
$\widetilde{\mu}(\omega)$, regardless of the nature of the heat
bath.  In particular, it is independent of the external potential
and the temperature \cite{ford}.  Further details may be found in
\cite{roc}. 

\subsection{Resistively shunted Josephson junctions}

Our purpose here is to show how some of the principal well-known
theoretical results follow simply from the quantum Langevin
approach and also how they may be generalized.  This will also
provide an example of our statement that "-- the operator $x$ is
the quantum Langevin equation -- can be a generalized displacement
operator --".  For an ideal junction the current is given by the
Josephson equation \cite{barone}:
\begin{equation} I=I_{C}\sin\phi,
\end{equation} where $\phi$ is the phase difference of the
superconducting wave function across the junction and $I_{C}$ is
the critical current.  The voltage across the junction is
\begin{equation} V=\frac{\hbar}{2e}\dot{\phi},
\end{equation} where $\frac{\hbar}{2e}$ is the quantum of flux.  A
real junction can be viewed as a capacitance $C$ and a shunt
resistance $R$ in parallel with an ideal junction.  The current is
then the sum of the ideal junction current, given by (3.19), the
current through the capacitor, $\dot{Q}= C\dot{V}$, and the
current through the resistor, $I=\frac{V}{R}$.  The junction
voltage is still given by (3.20).  The basic equation of motion of
the junction can therefore be written
\begin{equation}
\left(\frac{\hbar}{2e}\right)^{2}C\ddot{\phi}+\left(\frac{\hbar}{2e}\right)^{2}\frac{\dot{\phi}}{R}+\frac
{\hbar}{2e}I_{C}\sin\phi=\frac{\hbar}{2e}I+F(t).
\end{equation}  This is of the form of a quantum Langevin equation
with mass and friction constant
\begin{equation}
m=\left(\frac{\hbar}{2e}\right)^{2}C,~~~\zeta=\frac{(\frac{\hbar}{2e})^{2}}{R},
\end{equation} and with potential
\begin{eqnarray} U(\phi)=-\frac{\hbar}{2e}(I\phi+I_{C}\cos\phi).
\end{eqnarray}  Thus, analogous to (3.17), the ensemble average of
the square of the phase due to noise is
\begin{equation}
<\phi^{2}>=\frac{4e^{2}}{\pi{C}\hbar}\int_{0}^{\infty}d\omega\coth\left(\frac{\hbar\omega}{2kT}\right)
\frac{\omega\gamma}{(\omega^{2}_{0}-\omega^{2})^{2}+\omega^{2}\gamma^{2}},
\end{equation} where
\begin{equation}
\gamma=\frac{1}{RC},
~~~\omega^{2}_{0}=\frac{2e}{C\hbar}(I^{2}_{C}-I^{2})^\frac{1}{2}.
\end{equation}  In the limit of large shunt resistance (weak
coupling limit), $\gamma\ll\omega_{0}$, this becomes
\begin{equation}
<\phi^{2}>=\frac{2e^{2}}{C\hbar\omega_{0}}\coth\left(\frac{\hbar\omega_{0}}{2kT}\right).
\end{equation}  This weak coupling limit corresponds to the
expression for the phase fluctuations obtained by Josephson
\cite{josephson}.  The power spectrum of the voltage fluctuations
is readily  obtained from (3.24) using (3.20).  Further details
may be found in \cite{ford1988}.

\subsection{Environmental Effects on Nanosystems}

The study of nanosystems is now burgeoning and, because of the
small dimensions involved, environmental effects play an important
role.  To illustrate, we will focus our discussion on the analysis
of charge fluctuations on small-capacitance tunnel junctions. 
Following Ingold and Nazarov \cite{ingold}, we consider a junction
of capacitance $C$ carrying the charge
$Q=CV$ where $V$ is the voltage across the junction.  The external
circuit (the environment) is described by its impedance
$Z(\omega)=\frac{V(\omega)}{I(\omega)}$, where
$I(\omega)$ is the current.  The phase difference across the
junction is obtained from (3.20) [except the $2e\rightarrow e$
since here we are not considering a superconducting tunnel
junction] to get
\begin{equation}
\phi(t)=\frac{e}{\hbar}\int_{-\infty}^{t}dt^{\prime}V(t^{\prime}).  
\end{equation}  Turning to a quantum picture, it follows that
$(\frac{\hbar}{e})\phi$ and
$Q$ are the electrical equivalent of the mechanical quantities $x$
and $p$ (see table I of Ref. 35 for more details).  It readily
follows that there is a fluctuation-dissipation theorem describing
the relation between the charge fluctuation $q(t)$ and the
dissipation described by $Z(\omega)$.  Explicitly, we have
\cite{hu}
\begin{equation}
<q^{2}(t)>=\int^{\infty}_{0}d\omega\frac{\hbar\omega{C}^{2}}{\pi}\coth\frac{\hbar\omega}{2k_{B}
T}Re\left[\frac{1}{i\omega{C}+Z^{-1}(\omega)}\right].
\end{equation}  This result is applicable to any kind of external
circuit attached to a small junction.  In particular, this result
was used to calculate the effect of quantum smearing of Coulomb
blockade in small tunnel junctions \cite{hu94} and it was shown
that, in the weak coupling limit, the QLE theory is related to the
well-known generalized Landauer formula \cite{landauer}.  Finally,
we note that big efforts are currently being made toward
developing single charge transfer devices and, in that context, we
have used (3.27) to calculate environmental efforts on a single
electron box \cite{hu95}.

\subsection{Quantum Transport for a many-body System}

The problem of electrical conductivity in solids has been studied
by a large number of theoretical methods.  For the most part, they
are either based on the Boltzmann transport equation for the
distribution of electrons, or they start with the Kubo formula in
linear-response theory.  In spite of their successes, those two
methods are known to be impractical tools for calculating the
electrical transport properties of systems with high-order
impurity scattering, among other disadvantages.  Thus, we were
motivated to treat the quantum transport of an interacting system
of electrons, impurities, and phonons, in a time-dependent
electric field, by using the QLE in which the system is shown to
be equivalent to a quantum particle in a heat bath \cite{hu87}. 
The center of mass of the electrons acts like a quantum particle,
while the relative electrons and phonons play the role of a heat
bath.  They are coupled through the electron-impurity and
electron-phonon interactions.  After eliminating the heat-bath
variables, the equation of motion for the quantum particle is
written in a form of a QLE, with a memory term which reflects the
retarded effects of the heat bath on the quantum particle.  The
evaluation of the memory term immediately leads to a result for
the susceptibility from which we can calculate the conductivity
directly, in contrast to Kubo-type calculations which require the
evaluation of correlation functions as an intermediate step.  Our
results were then applied successfully to transport in 3, 2, 1,
and 0 dimensional systems and we refer to our review \cite{roc91}
for details.  As emphasized in \cite{roc91}, a big advantage of
the QLE approach is that it gives a natural separation between the
conductivity and the noise.

\subsection{Unruh radiation}

It is generally accepted that a system that undergoes uniform
acceleration with respect to the vacuum of flat-space-time will
thermalize at a temperature (the so-called Unruh temperature) that
is proportional to the acceleration \cite{unruh}.  However, the
question of whether or not the system actually radiates is highly
controversial.  Thus, we were motivated to carry out an exact
analysis of the problem using a generalized quantum Langevin
equation to describe the motion of an oscillator (the detector)
moving under a constant force and coupled to a one-dimensional
scalar field (scalar electrodynamics).  We conclude that the
system does not radiate despite the fact that it does in fact
thermalize at the Unruh temperature \cite{ford0}.  The essence of
the calculation was to show that the equation of motion of the
oscillator has the same form as that given in (3.3) except that
now the autocorrelation given in (2.5) not only corresponds to an
Ohmic heat bath but the relevant temperature is the Unruh
temperature which, of course, is proportional to the
acceleration.  However, due to the fact that there is no term
analogous to the external force $f(t)$ appearing (2.2), it
immediately follows that there is no radiation, as confirmed by a
detailed calculation.  What we found is that the uniform
acceleration does not give rise to a constant force in the
equation of motion but, instead, it only appears explicitly in the
autocorrelation of the fluctuation force.  In other words, in this
case, the constant external acceleration does not cause radiation
but simply accentuates the noise.

\section{BLACKBODY RADIATION HEAT BATH MODEL AND APPLICATIONS}

The relevant Hamilton \emph{H} is the universally accepted
\emph{H} of quantum electrodynamics (QED) from which it follows
\cite{ford},\cite{ford85} that
\begin{equation}
Re[\tilde{\mu}(\omega+i0^{+})]=\frac{2e^{2}\omega^{2}}{3c^{3}}f^{2}_{k},
\end{equation} where the quantity $f_{k}$ is the electron form
factor (Fourier transform of the electron charge distribution). 
In other words, we have allowing the electron to have structure.

The physically significant results for this model should not
depend upon details of the electron form factor, subject, of
course, to the condition that is be unity up to some large
frequency $\Omega$ and falls to zero thereafter.  A convenient
form which satisfies this condition is 
\begin{equation}
f^{2}_{k}=\frac{\Omega^{2}}{\omega^{2}+\Omega^{2}}.
\end{equation}  Using this in (4.1), the Stieltjes inversion
formula gives
\begin{equation}
\tilde{\mu}(z)=\frac{2e^{2}\Omega^{2}}{3c^{3}}\frac{z}{z+i\Omega}.
\end{equation}  We see here a manifestation of the general feature
that the memory function is independent of the external potential
and the particle mass.

As essential aspect of QED theory is the necessity for mass
renormalization.  Thus the
$m$ occurring in the QLE is actually the bare mass and the
renormalized (observed) mass
$M$ is given in terms of the bare mass $m$ by the relation
\begin{equation}
M=\frac{m+2e^{2}\Omega}{3c^{3}}=m+\tau_{e}\Omega{M},
\end{equation} where
\begin{equation}
\tau_{e}=\frac{2e^{2}}{3Mc^{3}}\simeq{6}\times{10}^{-24}\sec.
\end{equation}  Next, taking the inverse Fourier transform of
(4.3) leads to the result
\cite{ford91} 
\begin{equation}
\mu(t)=M\Omega^{2}\tau_{e}[2\delta(t)-\Omega\exp(-\Omega{t})],
\end{equation}  which makes manifest the non-Markovian nature of
the motion.  We now consider various applications.

\subsection{Elimination of Runaway Solutions in Electron Radiation
Theory}

The fact that $\tilde{\mu}(\omega)$ is a positive function (which
follows from the second law of thermodynamics) is equivalent to
the demand that all the poles of
$\alpha(\omega)$, the generalized susceptibility, must lie in the
lower half of the complex plane (which also follows from the
principle of causality).  This implies that
$m\ge0$ and that
\begin{equation}
\Omega\le\tau^{-1}_{e}=1.6\times{10}^{23}~~s^{-1}.
\end{equation}  If one selects for $\Omega$ its largest
permissible value: namely
$\tau^{-1}_{e}$ ( corresponding to letting $m\rightarrow{0}$,
which is also equivalent to choosing the closest approach to a
point electron consistent with causality), then one obtains the
classical equation of motion
\begin{equation} M\ddot{x}=f(t)+\tau_{e}\dot{f}(t).
\end{equation}  This result has a variety of desirable properties
not exhibited by the Abraham-Lorentz equation.  Generalizations
which include quantum effects and the presence of a potential may
be found in \cite{ford91},\cite{ford98}.  The Larmor formula also
requires generalization \cite{ford1991},\cite{fordroc}.  A
relativistic generalization of (4.8) is given in \cite{fordroc93}
along with a proof that a constant force does not cause radiation.

\subsection{Rydberg Atom Level Shifts due to Blackbody Radiation}

Atoms coupled to a bath in thermal equilibrium at temperature $T$
are best described in terms of the free energy \cite{ford85} and
we have argued \cite{ford86},\cite{fordlew87} that this is the
quantity actually measured by experiment \cite{hollberg}.  Thus,
using (2.19), we found that, in the high temperature limit, where
$kT$ is large compared with the level spacing, the shift in free
energy is given by 
\begin{equation}
\Delta{F}=\frac{\pi{e}^{2}(kT)^{2}}{9\hbar{M}c^{3}},
\end{equation}  in agreement with the experimental results
\cite{hollberg}.

\section{DECOHERENCE}

The superposition principle (and related work on Schr\"{o}dinger
cats), entanglement, and the quantum-classical interface are at
the cutting-edge of topical research
\cite{ford02}. Since superposition states are very sensitive to
decoherence, reservoir theory has attracted much recent interest.
Recognizing that conventional master equation approaches are often
not adequate, some investigators have used path integral methods. 
However, we have found that the simplest and most physically
appealing approach to the problem is via use of the QLE,
supplemented by use of the Wigner distribution function (WDF) for
the study of a system with an infinite spectrum of states
\cite{roc03} and by use of the spin polarization vector in the
case of spin systems \cite{roc05}.  For the investigation of
decoherence phenomena, we have found that it is important to
distinguish between two different physical scenarios: namely (a)
complete entanglement between the quantum particle and the heat
bath at all times \cite{fordlew01} and (b) the system in a state
in which the oscillator is not coupled to the bath at, say, $t=0$
and such that the bath is in equilibrium at temperature
$T$ \cite{fordroc01}.  Thus, it takes a characteristic time of the
order of
$\gamma^{-1}$ (where $\gamma$ is a typical dissipative decay time)
for the complete coupling to occur and for the whole system to
come into thermal equilibrium.  It is noteworthy that "decoherence
without dissipation "can occur [55-57] but not for low
temperatures \cite{fordroc03}.

Scenario (a), the "entanglement at all times" calculation utilized
quantum probability distributions (which are related to Wigner
distributions) in conjunction with results obtained by use of the
stationery solution to the QLE, given in (2.2).  On the other
hand, for the scenario (b) case, it is necessary to take into
account initial conditions.  The Langevin equation for the
oscillator with given initial values is given by
\cite{ford87},\cite{fordroc01}
\begin{equation}
m\ddot{x}+\int^{t}_{0}dt^{\prime}\mu(t-t^{\prime})\dot{x}(t^{\prime})+Kx=-\mu(t)x(0)+F(t),
\label{qle9}
\end{equation} and the general solution is given by
\begin{equation} x(t)=m\dot{G}(t)x(0)+mG(t)\dot{x}(0)+X(t),
\label{qle11}
\end{equation} where we have introduced the fluctuating position
operator,
\begin{equation}
X(t)=\int^{t}_{0}dt^{\prime}G(t-t^{\prime})F(t^{\prime}).
\label{qle39}
\end{equation}

If we assume that at $t=0$ the system is in a state in which the
oscillator is not coupled to the bath and that the bath is in
equilibrium at temperature $T$, we find that the correlation and
commutator are the same as those for the stationary equation.  It
was then possible to show \cite{fordroc01} that one could write
the Langevin equation (\ref{qle9}) in the form of an equation that
is local in time with time-dependent coefficients:
\begin{equation}
\ddot{x}+2\Gamma(t)\dot{x}+\Omega^{2}(t)x=\frac{1}{m}F(t),
\label{qle42}
\end{equation} where explicit expressions for $\Gamma (t)$ and
$\Omega (t)$ were obtained in terms of $G(t)$.  Furthermore, it
was shown that these results constitute in essence a derivation of
the HPZ exact master equation
\cite{hu92} with explicit expressions for the time-dependent
coefficients.  It is also notable that in this scenario one needs
to assume that the initial temperature of the particle is the same
as that of the heat bath in order to obtain "decoherence without
dissipation".

\section{DRIVEN SYSTEMS}

As remarked in Sec. II, the position autocorrelation, as distinct
from the fluctuation force autocorrelation, is modified by a
presence of an external force $f(t)$, as discussed in detail in
\cite{fordroc00},\cite{fordroc2000}.  It is convenient to now
write 
\begin{equation} x(t)=x_{s}(t)+x_{d}, \label{qle64}
\end{equation}and
\begin{equation} C(t)=C_{0}(t)+C_{d}(t),
\end{equation} where $x_{d}$ is the "driven" contribution due to
the external force $f(t)$ and $x_{s}$ is the contribution due to
the fluctuation force $F(t)$.  Here we have introduced a subscript
s to emphasize that
$x_{{\rm s}}(t)$ is a stationary operator-process, in the sense
that correlations, probability distributions, etc. for this
dynamical variable are invariant under time-translation
($t\rightarrow t+t_{0}$). In particular, the correlation
$C_{0}(t)$ is given by the right-side of (2.14) and  is a function
only of the time-difference
$t-t^{\prime }$.  Furthermore, \cite{fordroc2000}
\begin{eqnarray} C_{d}(t,t^{\prime})=\langle x(t)\rangle \langle
x(t^{\prime})\rangle, \label{qle65}
\end{eqnarray} where $\langle x(t)\rangle$ is the steady mean of
the driven motion.  Based on these results, we extended the
calculation of the well-known Burshtein-Mollow spectrum of
resonance fluorescence to the case of non-zero temperature.  In
the high temperature limit, $\kappa{T}>>\hbar\omega_{0}$, where
$\hbar\omega_{0}$ is the resonant energy for the two-level (atom),
we found that the decay rate is increased by a factor
$\left(\frac{\kappa{T}}{\hbar\omega_{0}}\right)$.  Our calculation
was based on use of the Lax formula for two-time correlations but,
as we also pointed out, this formula is applicable only for weak
coupling and for frequencies near a resonance frequency since,
more generally, the Onsager classical regression theorem cannot be
generalized to the quantum domain
\cite{fordroc2000},\cite{fordroc96}.

A further application of these results was the consideration of
decoherence phenomena in the presence of an external field
\cite{roczuo},\cite{zuo}.  This work was motivated by the recent
interest in engineered reservoirs \cite{schleich}.  In particular,
Myatt et al. \cite{myatt} used a linear Paul trap to confine
single Be ions in a harmonic potential and then prepared various
superposition states.  Next, they induced decoherence by coupling
the single ion to a reservoir which they controlled in various
ways.  Such a reservoir gives rise to an external force
$f(t)$ in the equation of motion of the system.  The calculations
in
\cite{roczuo},\cite{zuo} led to the conclusion that a non-random
external field does not give rise to decoherence whereas, by
contrast, a random field does.  In particular, the experiments of
Myatt et al. \cite{myatt} used a $\delta$-correlated force.  The
existing experiments verified the familiar result that the
decoherence decay time $\tau_{d}$ is inversely proportional to the
square of the separation $d$ of the superposition components. 
However, this is a familiar result predicted by the plethora of
papers dealing with the
$f(t)=0$ situation but it does not give information on the
dependence of
$\tau_{d}$ on the parameters of the externally-superimposed
reservoir.  Thus, we will have to await come experimental data in
order to compare with existing theory.

\section{Acknowledgement}

The author is pleased to acknowledge that all of the essential
results described above were derived in collaboration with Prof.
G. W. Ford.

\end{document}